\documentclass[aps,eqsecnum,nofootinbib,superscriptaddress,showpacs,preprintnumbers]{revtex4}
\usepackage{amssymb, amsmath, amsopn, amsthm,mathptmx}
\usepackage{epsfig,amsfonts,bm}
\usepackage{MnSymbol,color}
\usepackage{lscape}
\usepackage{graphicx}
\def\eref#1{Eq.\ \eqref{#1}}

\newcommand{\nt}{\notag\\}

\begin{document}
\preprint{KOBE-COSMO-19-07}

\title{Search for ultralight scalar dark matter with NANOGrav pulsar timing arrays}
\date{\today}
\begin{abstract}
An ultralight scalar field is a candidate for the dark matter.
The ultralight scalar dark matter with mass around $10^{-23}\,{\rm eV}$ induces oscillations of the pulse arrival time in the sensitive frequency range of the pulsar timing arrays.
We search for the ultralight scalar dark matter using the North American Nanohertz Observatory for Gravitational Waves 11-year Data Set.
We give the 95\% confidence upper limit for the signal induced by the ultralight scalar dark matter.
In comparison with the published Bayesian upper limits on the amplitude of the ultralight scalar dark matter obtained by Bayesian analysis using the Parkes Pulsar Timing Array 12-year data set (Porayko et al. 2018), we find three times stronger upper limit in the frequency 
range from $10^{-8.34}$ to $10^{-8.19}\,{ \rm Hz}$ which corresponds to the mass range from $9.45\times10^{-24}$ to $1.34\times10^{-23}\,{\rm eV}$. 
In terms of the energy density of the dark matter, we find that  the energy density near the Earth is less than $7\,{\rm GeV/cm^3}$ in the range from $10^{-8.55}$ to $10^{-8.01}\,{ \rm Hz}$ (from $5.83\times10^{-24}$ to $2.02\times10^{-23}\,{\rm eV}$).
The strongest upper limit on the the energy density is given by $2\,{\rm GeV/cm^3}$ at a frequency  $10^{-8.28}\,{ \rm Hz}$ (corresponding to a mass $1.09\times10^{-23}\,{\rm eV}$).
We find that the signal of the ultralight scalar dark matter can be explained by the solar system ephemeris effect.
Also, we reveal that the model of the solar system ephemeris effect prefers parameters which are contrary to the expectation that noise will be reduced on all pulsars.
\end{abstract}
\author{Ryo Kato}
\email{153s107s@stu.kobe-u.ac.jp}
\author{Jiro Soda}
\email{jiro@phys.sci.kobe-u.ac.jp}
\affiliation{Department of Physics, Kobe University, Rokkodai 1-1, Kobe 657-8501, Japan}

%\pacs{?}

\maketitle

\section{Introduction}\label{sec:introduction}

The dark matter problem is clearly one  of the most important issues in modern cosmology.
Recently, motivated by string theory, an ultralight scalar dark matter has been intensively studied~\cite{Svrcek:2006yi,Arvanitaki:2009fg}.
In particular, an ultralight scalar field with mass $10^{-23} \,{\rm eV}$ can behave like the cold dark matter (CDM) on cosmological scales 
and resolve a cusp problem~\cite{PhysRevLett.85.1158,Marsh:2015xka}. 
In this article, we call it simply the fuzzy dark matter (FDM).
The FDM can be treated as a classical scalar field because the occupation number of the FDM accounting for the energy density of the dark matter is very large.
The main difference between FDM and CDM is that  the pressure of the FDM is coherently oscillating, while that of CDM almost vanishes.
Khmelnitsky and Rubakov have pointed out that the effect of oscillating pressure might be detected with the pulsar timing arrays (PTAs)~\cite{Khmelnitsky_2014}.
Indeed, the oscillation of the pressure induces the oscillation of the gravitational potential, and as a result, it induces the oscillation of the arrival time of the pulse passing through the gravitational potential.

It would be worth noting that there exists experimental duality between gravitational wave and scalar dark matter detections. 
More precisely, the detection method for gravitational waves is useful for scalar field dark matter, and vice versa.
The idea of Khmelnitsky and Rubakov inspired us to use the gravitational wave interferometers for detecting scalar dark matter~\cite{Aoki:2016kwl}.
Recently, the importance of the reverse direction has been promoted and a novel constraint on GHz gravitational waves was obtained~\cite{Ito:2019wcb}.
Hence, it is importnt to  investigate thoroughly the duality.

An attempt to search for long wavelength gravitational waves with the PTAs composed of long-term observation of many pulsars was proposed in the articles \cite{1979ApJ...234.1100D,Romani_1989,1990ApJ...361..300F}.
Nowadays, the PTAs are most sensitive to the gravitational waves with a few nanohertz frequency.
There are three major pulsar timing projects aimed at observing the pulsars and searching for the gravitational waves: the European Pulsar Timing Array (EPTA) \cite{Kramer_2013}, the North American Nanohertz Observatory for Gravitational Waves (NANOGrav) \cite{McLaughlin_2013}, and the Parkes Pulsar Timing Array (PPTA) \cite{manchester_2013}.
The collaboration of the three projects is called the International Pulsar Timing Array (IPTA) \cite{Hobbs_2010}.
The point is that we can utilize the PTAs for searching the FDM.

Porayko and Postnov \cite{PhysRevD.90.062008} gave upper limits for the FDM with the Bayesian analysis using the NANOGrav 5-year Data Set.
Moreover, Porayko et al. \cite{Porayko:2018sfa} gave upper limits for the FDM with the Bayesian and the Frequentist analyses using the PPTA 12-year Data Set.
In this article, following the previous articles, we search for the FDM by the Bayesian analysis in the time domain using the NANOGrav 11-year Data Set.
We quantitatively investigate whether the ultralight scalar dark matter is detectable or not using the Bayesian model selection approach.
We clarify the prior dependence of constraints on the amplitude of the FDM and obtain three times stronger constraints on the amplitude of the FDM in the frequency 
range from $10^{-8.34}$ to $10^{-8.19}\,{ \rm Hz}$.  
We also discuss how the results of Bayesian analysis depend on the the solar system ephemeris noise in the model describing the observation data.

This article is organized as follows. In Section \ref{sec:FDM Signal} we describe a model of FDM signal.
In Section \ref{sec:Bayesian Parameter Estimation and Model Comparison} we briefly review the Bayesian statistics, and explain how to use it for our analysis.
In Section \ref{sec:Bayesian analysis in the time domain} we describe the model, the data, and the function used in the Bayesian analysis. 
In Section \ref{sec:Markov chain Monte Carlo simulation} we briefly review the MCMC simulation.
In Section \ref{sec:Pre-analysis} we describe the analysis of the white noise that is performed before the main analysis.
In Section \ref{sec:Results} we summarize the results of Bayesian analysis using NANOGrav 11-year data set.
The last Section is devoted to conclusion.

\section{FDM Signal}\label{sec:FDM Signal}
As we mentioned in the introduction, the oscillation of the scalar field with the mass $m$ induces the oscillation of the gravitational potential
and the oscillation of the arrival time of the pulse passing through the gravitational potential.
The oscillation of the arrival time of the pulse induced by the FDM is given by~\cite{Khmelnitsky:2013lxt}
\begin{eqnarray}
s(t)=-\frac{1}{2\pi f}[\Psi({\bf x_{e}})\sin(2\pi f t+\alpha({\bf x_{e}}))-\Psi({\bf x_{p}})\sin(2\pi f (t-D)+\alpha({\bf x_{p}}))],
\end{eqnarray}
where $f=m/\pi$ is the frequency, $m$ is the mass, $D$ is the distance between the pulsar and the earth, each ${\bf x_{e}}$ and ${\bf x_{p}}$ are the position of the Earth and the pulsar, and $\alpha$ denotes the phase.
Here, we used the gravitational potential
\begin{eqnarray}
\quad \Psi({\bf x})\equiv\frac{\pi G \rho({\bf x})}{m^2},     \label{psi_f}
\end{eqnarray}
where  $\rho({\bf x})$ is the energy density of the dark matter.
We will refer to this oscillation as the FDM signal.

The parameters used in the Bayesian estimation are defined as follows:
\begin{eqnarray}
s(t)=-\frac{\Psi}{2\pi f}[\sin(2\pi f t+\alpha_{e})-\sin(2\pi f t+\alpha_{p})],\label{axion_signal}
\end{eqnarray}
where we assumed $\Psi\equiv\Psi({\bf x_{e}})=\Psi({\bf x_{p}})$ and defined
\begin{eqnarray}
\alpha_{e}\equiv\alpha({\bf x_{e}}),\quad \alpha_{p}\equiv\alpha({\bf x_{p}})-2\pi f L \ .
\end{eqnarray}
Here, since we do not aim to estimate the distance, we put together the phase  $\alpha({\bf x_{p}})$ and the distance  $L$.
In fact, the distance has an uncertainty of tens to hundreds of parsec currently that is too large to determine the phase \cite{Arzoumanian:2017puf}.
Since, as is mentioned in the article \cite{Porayko:2018sfa}, the distance between the Earth and the pulsar $L$ is not so large, it is reasonable to assume that the amplitudes at the earth $\Psi({\bf x_{e}})$ and the pulsar $\Psi({\bf x_{p}})$ are equal.
The FDM signal is a simple sine wave superposition, and so some noise in the data may have a similar waveform.
Therefore, it is necessary to consider whether the FDM signal obtained by data analysis is the actual FDM signal or some noise.

If all dark matter is occupied by the FDM, the predicted amplitude at the position of the Earth is
\begin{eqnarray}
\Psi&\simeq&6.48\times10^{-16}\left(\frac{\rho}{0.4 {\rm Gev/cm^3}}\right)\left(\frac{10^{-23}{\rm eV}}{m}\right)^2,\nt
&\simeq&9.47\times10^{-16}\left(\frac{\rho}{0.4 {\rm Gev/cm^3}}\right)\left(\frac{4\times10^{-9}{\rm Hz}}{f}\right)^2,     \label{Psi}
\end{eqnarray}
and
\begin{eqnarray}
f \simeq4.84\times10^{-9} {\rm Hz} \left(\frac{m}{10^{-23}{\rm eV}}\right),        \label{f}
\end{eqnarray}
where $\rho=0.4 {\rm Gev/cm^3}$ is the estimated energy density of the dark matter at the position of the Earth \cite{Nesti_2013}.

\section{Bayesian Parameter Estimation and Model Comparison}\label{sec:Bayesian Parameter Estimation and Model Comparison}
In this section we explain the Bayesian parameter estimation and the model comparison. For further details about the Bayesian data analysis, see for example \cite{gregory_2005, kruschke2010doing,gelman2013bayesian}.

The purpose of the Bayesian parameter estimation is to estimate the posterior probability distribution $p(\theta|D)$ of the parameters $\theta$ given the data $D$.
Having the data, we can update our belief about the parameters using Bayes' rule, namely
\begin{eqnarray}
p(\theta|D)=\frac{p(D|\theta)p(\theta)}{p(D)},\label{bayes theorem}
\end{eqnarray}
where $p(\theta)$ is the prior probability distribution, $p(D|\theta)$ is the likelihood function, and $p(D)$ is the evidence. 
In the parameter estimation, the evidence can be regarded as a normalization constant which can be omitted, because it does not involve the parameter.
We will define the data $D$, the parameter $\theta$, the likelihood function $p(D|\theta)$, and the prior probability distribution $p(\theta)$ in Section \ref{sec:Bayesian analysis in the time domain}.

Using the posterior probability distribution for the amplitude of the FDM signal $\Psi$, we define the 95\% upper limit $R$ by
\begin{eqnarray}
\int_{0}^{R}p(\Psi|D)d\Psi=0.95.\label{upper limit}
\end{eqnarray}
We will calculate this 95\% upper limit with the samples of the posterior probability distribution which is generated by the MCMC in Section \ref{subsec:Upper limits}.

\begin{table}[t]
\caption{Interpretation of the Bayes factor}
\label{jeffreys scale}
\begin{tabular}{llll}\hline
$B_{12}$ \qquad\quad& Evidence in favor of $\mathcal{M}_{1}$ against $\mathcal{M}_{2}$ \\ \hline
1-3  & Not worth more than a bare mention \\
3-20  & Positive \\
20-150 & Strong \\
$>$150 & Very strong \\ \hline
\end{tabular}
\end{table}

For the Bayesian model comparison, it is often considered the Bayes factor:
\begin{eqnarray}
B_{12}=\frac{p(D|\mathcal{M}_{1})}{p(D|\mathcal{M}_{2})},
\end{eqnarray}
where $\mathcal{M}_{1}$ and $\mathcal{M}_{2}$ are competing models which assign a meaning to the parameters.
The Table I gives the interpretation of the Bayes factor in terms of the strength of the evidence \cite{jeffreys1948theory,doi:10.1080/01621459.1995.10476572,10.2307/271063}.

If $\mathcal{M}_{1}$ and $\mathcal{M}_{2}$ are nested models, we can use the Savege-Dickey density ratio to calculate the Bayes factor \cite{o1994kendall,10.2307/2958475,doi:10.1080/01621459.1995.10476554,doi:10.1080/01621459.1995.10476572}.
We assume that $\mathcal{M}_{1}$ is a model described in Section \ref{sec:Model}, and $\mathcal{M}_{2}$ is a model in which the amplitude $\Psi$ of the model $\mathcal{M}_{1}$ is fixed to a very small value $10^{-18}$.
In this case, the model $\mathcal{M}_{2}$ can be regarded as a model with no FDM signal.
Then the Savege-Dickey density ratio is
\begin{eqnarray}
B_{12}=\frac{p(\Psi=10^{-18}|\mathcal{M}_{1})}{p(\Psi=10^{-18}|D,\mathcal{M}_{1})}.\label{SDDR}
\end{eqnarray}
We will calculate this Bayes factor with the samples of the posterior probability distribution which is generated by the MCMC in Section \ref{sec:Bayesian analysis in the time domain}.
Specifically, we calculate the Bayes factor in multiple small bins around a fixed value $\Psi_{0}$, then derive the mean and unbiased standard deviation.

\section{Bayesian analysis in the time domain}\label{sec:Bayesian analysis in the time domain}

In this section, we explain the data  $D$, the model $\mathcal{M}$, and the parameter $\theta$, and define the likelihood function $p(D|\theta)$, and the prior probability distribution $p(\theta)$.

\subsection{Data}\label{sec:Data}
We used the NANOGrav 11-year data set~\cite{Arzoumanian:2017puf} and chose six pulsars: PSRs J0613-0200, J1012+5307, J1600-3053, J1713+0747, J1744-1134, and J1909-3744.
In this dataset, these pulsars have relatively good time-of-arrival (TOA) precision and long observation period, which would be suitable for detecting the FDM signal which becomes larger as the frequency becomes lower.

The data $D$ for the Bayesian analysis are timing residuals which are calculated by subtracting the timing model from the TOAs \cite{Edwards:2006zg}.
The fitting to make timing residual is called timing fit, which removes the currently well-known effects.
The parameters included in the timing model are spin parameters, astrometry parameters, binary parameters (if pulser is binary), dispersion measure parameters, frequency dependency parameters, jump parameters, see the articles \cite{2015ApJ...813...65T,Arzoumanian:2017puf} for details.
Note that, by fitting the spin parameters, some of the low frequency signal that we are searching for in this article will be absorbed \cite{1984JApA....5..369B,Cutler:2013aja}.
In the timing fit, TT (BIPM2015)\footnote{https://www.bipm.org/en/bipm-services/timescales/time-ftp/ttbipm.html} is used for the Terrestrial Time, and JPL DE436 \cite{Folkner2016} is used for the planetary ephemeris.
It is known that the timing residuals change greatly depending on which planetary ephemeris is selected.
To account for this error, the SSE noise described in Section \ref{sec:Model} was first introduced to the model by the article \cite{Arzoumanian:2018saf}.

In order to obtain the timing residuals, we use the libstempo\footnote{http://vallis.github.io/libstempo} which is the PYTHON interface to TEMPO2\footnote{https://bitbucket.org/psrsoft/tempo2.git} \cite{doi:10.1111/j.1365-2966.2006.10302.x} timing package.\footnote{We confirmed that each pulsar's value of the chi-square and the degrees of freedom which can be derived by TEMPO2 are consistent with values listed in the file `stats\_11y\_20180226.dat', where this file is included in the data set and can be used as to see if TEMPO2 is properly constructed. Therefore, TEMPO2 was installed as expected.}
We used the identical data set except for the parameter file of PSR J1713+0747.
In the parameter file of the PSR J1713+0747, we changed only a parameter EPHEM from DE430 \cite{2014IPNPR.196C...1F} to DE436, where this parameter specifies which ephemerides to be used.
Then we used libstempo to fit the timing parameters of the PSR J1713+0747 and created a new parameter file.
We iterated the timing fitting five times, which would be sufficient for parameters to converge to certain values.
We found that the LAMBDA and BETA parameters of the astrometry parameters shift by more than one sigma away from the older value.
This result can be inferred from the deviation of the position parameter due to the difference in ephemeris as pointed out in the article \cite{Wang:2017lth}.

\subsection{Model}\label{sec:Model}
Following the paper \cite{2015ApJ...813...65T,Arzoumanian:2018saf}, the model $\mathcal{M}_{1}$ of the timing residuals $\delta \bm{t}$ for each pulsar includes the FDM signal, the linearized timing model noise, the red noise, the Solar System ephemeris noise, and the white noise.
Each noise is described below.

First, the linearized timing model noise characterize the inaccuracies of the timing fit.
The linearized timing model noise is represented by the product of a design matrix and small offsets.
The design matrix describes the dependence of the timing residuals on respective timing model parameters.
We obtain the design matrix using the TEMPO2 via libstempo.
The timing model parameters are listed in \cite{Arzoumanian:2017puf}. 

Second, the red noise has most of their power at low frequencies in a given data set.
The red noise is known to have achromatic (observing-frequency-independent) and chromatic (observing-frequency-dependent) components \cite{2017ApJ...834...35L}.
The achromatic components are thought to be caused by a random walk in one of the pulsar spin parameters \cite{1972ApJ...175..217B,1975ApJS...29..453G,0004-637X-725-2-1607,Lyne408} and contributions to TOAs by an asteroid belt around the pulsar \cite{2013ApJ...766....5S}.
The chromatic components are thought to be caused by the pulse propagating through the ionized interstellar medium \cite{2016ApJ...819..155L,2017ApJ...834...35L}.
Although the origins of red noise are various, simple power-law spectrum form is often used as the power spectral density.
Specifically, the power spectral density of the red noise $P(f)$ is defined as
\begin{eqnarray}
P(f) =  \frac{A^2}{12\pi^2}\left(\frac{f}{f_{\rm yr}}\right)^{3-\gamma}f^{-3},\label{Pf}
\end{eqnarray}
where $f$ is a red noise frequency, $f_{\rm yr}$ is $1{\rm yr}^{-1}$, $A$ is a dimensionless amplitude of the red noise, and $\gamma$ is a spectral index of the red noise.

Third, the Solar System ephemeris noise characterizes the inaccuracy in the vector between the geocenter and the Solar System barycenter caused by the inaccuracies of the Solar System ephemeris.
We will refer to this noise as the SSE noise.
It is known that SSE noise affect upper limits and Bayes factors for amplitudes of the stochastic gravitational wave background~\cite{Arzoumanian:2018saf}.
The components of the SSE noise are considered to be the uncertainties of the planet mass and the planet orbit, and the uncertainty of the rotation rate around the ecliptic pole.
The uncertainties of the planet mass are taken into account in Jupiter, Saturn, Uranus and Neptune.
The uncertainties of the planet orbit are taken into account in Jupiter.
We used the values and the data implemented in ENTERPRISE (Enhanced Numerical Toolbox Enabling a Robust PulsaR Inference SuitE) which is a pulsar timing analysis code\footnote{https://github.com/nanograv/enterprise}.
Thus, regarding the SSE noise caused by the uncertainties of the planet orbit, the Jupiter mass is the value of the IAU 2009 system of astronomical constants \cite{Luzum2011}.
Also, the data describing the dependence of the Jupiter orbit on respective set-III parameters \cite{brouwer1961methods} is the same as ENTERPRISE.
Regarding the SSE noise caused by the uncertainty of the rotation rate around the ecliptic pole, the reference date corresponds to MJD 55197.

Finally, the white noise has the same power in all frequency region.
The components of the white noise are EFAC, EQUAD, and ECORR.
The EFAC characterize systematic errors of the TOA measurement uncertainties. 
The EQUAD is an additional white noise.
The EFAC and the EQUAD are used independently for each combination of receivers and backend systems.
The ECORR differs from the EQUAD in that there is a correlation between the TOAs obtained during a single observation.
This ECORR characterizes pulse jitter caused by stochastic amplitude and phase variations in pulse, which correlates in a certain frequency band and doesn't correlate in time \cite{2016ApJ...819..155L}.

\subsection{Likelihood Function}\label{sec:Likelihood Function and Posterior Probability Distribution}

The likelihood for each pulsar can be written as \cite{2015ApJ...813...65T, Taylor:2016gpq}:
\begin{eqnarray}
p(\delta \bm{t}|{\bm\theta})&=&\frac{1}{\sqrt{(2\pi)^{N_{\rm TOA}} \mathrm{det}({\bm C}_{\rm white})\mathrm{det}({\bm B})\det\left(\bm{T}^{T}{\bm C}_{\rm white}^{-1}\bm{T}+{\bm B}^{-1}\right)}}\nt
&&\times\mathrm{exp}\left(-\frac{1}{2}\left[\delta \bm{r}^{T}{\bm C}_{\rm white}^{-1}\delta \bm{r}-(\bm{T}^{T}{\bm C}_{\rm white}^{-1}\delta \bm{r})^{T}\left(\bm{T}^{T}{\bm C}_{\rm white}^{-1}\bm{T}+{\bm B}^{-1}\right)^{-1}\bm{T}^{T}{\bm C}_{\rm white}^{-1}\delta \bm{r}\right]\right),\label{marg likelihood}
\end{eqnarray}
where
\begin{eqnarray}
\delta \bm{r}&\equiv&\delta \bm{t} - \bm{s} - \bm{n}_{\rm SSE},\nt
\bm{T}&\equiv&(\bm{M}\,\, \bm{F}),\nt
\bm B&\equiv&\mathrm{diag}(\underbrace{10^{80},10^{80}, \cdots 10^{80}}_{N_{\rm TM}},\Xi_{1}, \Xi_{2}, \cdots \Xi_{2N_{\rm red}}), 
\label{DrTB}
\end{eqnarray}
$\delta \bm{t}$ is the $N_{\rm TOA}$ dimensional timing residuals, $N_{\rm TOA}$ denotes the number of TOAs of the pulsar,
$\bm s$ is the $N_{\rm TOA}$ dimensional FDM signal,
$\bm{n}_{\rm SSE}$ is the $N_{\rm TOA}$ dimensional linearized timing model noise,
$\bm M$ is a $N_{\rm TOA} \times N_{\rm TM}$ design matrix, $N_{\rm TM}$ denotes the number of the timing model parameters,
$\bm F$ is a $N_{\rm TOA} \times 2N_{\rm red}$ matrix which has columns of alternating cosine and sine functions for the red noise, $N_{\rm red}$  denotes the number of frequencies,
$10^{80}$ are the variance of the TM noise parameters,
$\Xi$ is the power spectrum of the corresponding frequency times the inverse of the observation period of the pulsar, 
and ${\bm C}_{\rm white}$ is the $N_{\rm TOA} \times N_{\rm TOA}$ covariance matrix which is composed by the EFAC, EQUAD, and ECORR.
The likelihood function using multiple pulsars is represented by the product of the likelihood function of each pulsar.

\subsection{Prior Probability Distribution}\label{sec:Prior Probability Distribution}
We use specific knowledge only for the mass errors of each planet as the prior information.
Using the propagation of uncertainty law, the variances of  Jupiter mass, Saturn mass, Neptune mass are calculated from the IAU 2009 system of astronomical constants.
Also, the variance of Uranus mass is calculated from the values in the article \cite{1538-3881-148-5-76} which is newer than the IAU 2009 system of astronomical constants.
Then we assume a normal distribution for the uncertainty of each planet mass and apply the obtained variances.
For parameters without specific knowledge, we use a log-uniform distribution for parameters which are need to be searched over several orders of magnitude with only positive values, and we use a uniform distribution for the other parameters.
The range of the log-uniform distribution and the uniform distribution is taken sufficiently wider than the value that the parameter would take.
Regarding the amplitude of the FDM signal, we especially consider both cases of uniform distribution and log-uniform distribution as in the article \cite{Arzoumanian:2018saf}.
The uniform distribution is used for the upper limit, and the log-uniform distribution is used for the model comparison.
The parameters and their prior probability distribution are given in Table \ref{Prior}.

As is done in \cite{0004-637X-794-2-141,0004-637X-821-1-13,Arzoumanian:2018saf}, we analyze the white noises in advance before the main analysis.
The resulting MCMC chains are used to calculate the value that maximizes the one-dimensional posterior probability distribution corresponding to each white noise, where this value is called the maximum a posteriori (MAP) value.
The main analysis is performed by fixing the possible values of the white noise parameter to the MAP value.
See the section \ref{sec:Pre-analysis} for the pre-analysis.

\begin{table}[t]
\caption{Prior Probability Distribution}
\label{Prior}
\begin{tabular}{llll}\hline
parameter\qquad&description\qquad&prior probability distribution \\
\hline
FDM signal\footnote{$f$ is the center frequency of the bin: $f=\{10^{-9},10^{-8.97},\cdots,10^{-8.01}\}$} &&  \\
$\Psi$ & amplitude & Uniform[$10^{-18}$,\,\,$10^{-11}$]\quad (for upper limit)\\
&&logUniform[$-18$,\,\,$-11$]\quad(for model comparison)\\
$f\quad[{\rm Hz}]$ & frequency & logUniform[$\log f-0.015$,\,\,$\log f+0.015$] \\
$\alpha_{e}\quad[{\rm rad}]$&phase at Earth & Uniform[$0$,\,\,$2\pi$] \\ 
$\alpha_{p}\quad[{\rm rad}]$&phase at pulsar & Uniform[$0$,\,\,$2\pi$] \\
\hline
red noise &&  \\
$A$ & amplitude & logUniform[$-20$,\,\,$-11$] \\
$\gamma$ & spectral index & Uniform[$0.02$,\,\,$6.98$] \\
\hline
SSE noise\footnote{$\mu$ is the subscript for the set-III parameters : $\mu=\{1,2,\cdots,6\}$} &&  \\
$\delta M^{\rm J}\quad[M_{\odot}]$ & mass error of Jupiter & $\mathcal{N}$($0$,\,\,$1.55\times{10^{-11}}$) \\
$\delta M^{\rm S}\quad[M_{\odot}]$ & mass error of Saturn & $\mathcal{N}$($0$,\,\,$8.17\times{10^{-12}}$) \\
$\delta M^{\rm U}\quad[M_{\odot}]$ & mass error of Uranus & $\mathcal{N}$($0$,\,\,$5.72\times{10^{-11}}$) \\
$\delta M^{\rm N}\quad[M_{\odot}]$ & mass error of Neptune & $\mathcal{N}$($0$,\,\,$7.96\times{10^{-11}}$) \\
$\delta a_{\mu}^{\rm J}$ & small offsets of set-III parameters & Uniform[$-0.05$,\,\,$0.05$] \\
$\delta z\quad[{\rm rad/year}]$ & rotation rate around ecliptic pole & Uniform[$-10^{-9}$,\,\,$10^{-9}$] \\
\hline
white noise &&  \\
EFAC & EFAC parameter & Uniform[$0.001$,\,\,$10$]\quad(for pre-analysis)\\
EQUAD$\quad[{\rm s}]$ & EQUAD parameter & logUniform[$-10$,\,\,$-4$] \quad(for pre-analysis)\\
ECORR$\quad[{\rm s}]$ & ECORR parameter & logUniform[$-8.5$,\,\,$-4$] \quad(for pre-analysis)\\
\hline
\end{tabular}
\end{table}

\section{Markov chain Monte Carlo simulation}\label{sec:Markov chain Monte Carlo simulation}
The MCMC simulation can be used to generate samples from the posterior probability distribution.
The MCMC method we used is called parallel tempering.
In the parallel tempering method, a concept of temperature is introduced, and the MCMC simulations of different temperatures are executed in parallel.
The advantage of parallel tempering is that it is possible to reduce the tendency of the samples of the posterior distribution to be trapped in a local minimum, compared to the Metropolis-Hastings method which is the one of the most famous MCMC methods \cite{gregory_2005}.
We carry out the analysis using four temperatures $T=1.00, 4.64, 21.5, 100$.

In order to perform the parallel tempering, we use the software package PTMCMCSampler\footnote{https://github.com/jellis18/PTMCMCSampler} \cite{justin_ellis_2017_1037579} via PAL2\footnote{https://github.com/jellis18/PAL2} \cite{justin_ellis_2017_251456} which is a Bayesian inference package for PTA and can include the PTMCMCSampler.
Regarding models not implemented in the PAL2, the FDM signal is implemented like the continuous gravitational waves and the SSE noise is implemented like any other noises.
Following the article \cite{0004-637X-794-2-141}, we use adaptive Metropolis \cite{10.2307/3318737}, single component adaptive Metropolis \cite{haario2005componentwise}, and differential evolution \cite{ter2006markov}, as a proposal algorithm which is used to generate next samples using past samples.
Furthermore, we also use a simple proposal algorithm to generate the next sample of each parameter by proposal distribution which is the same distribution as the probability distribution.
All of these proposal algorithms are used in a single MCMC simulation and which one is used is chosen randomly for each proposal in the MCMC simulation.
In this article, we use the value written in the PAL2 for each variable used in PTMCMCSampler, unless specifically mentioned.

\section{Pre-analysis}\label{sec:Pre-analysis}
As is usual~\cite{0004-637X-794-2-141,0004-637X-821-1-13,Arzoumanian:2018saf}, in order to obtain the MAP values of the parameters of the white noise, we analyze the white noise first before the main analysis.
By doing this, the number of the free parameters can be reduced in the main analysis.
In the pre-analysis, we performed independent analysis for each pulsar, and we used the model which contains the red noise in addition to white noise.
We ran the MCMC simulation with $10^{6}$ iterations and removed the first 25\% as a burn-in period, where the burn-in period is the period during which samples have not yet been obtained from the target distribution.

Table \ref{Result of pre-analysis} shows the results of the white noise and the red noise obtained by the pre-analysis.
The obtained posterior distribution is expressed by the MAP value and the 95\% confidence interval.
At the Green Bank Observatory, the receivers are Rcvr\_800 and Rcvr1\_2, and the backend system is GASP in early observations and GUPPI in later observations.
Similarly, at the Arecibo Observatory, the receivers are L-wide and S-wide, and the backend system is ASP in early observations and PUPPI in later observations.
Please see the article \cite{2015ApJ...813...65T} for details on the receiver and the backend system.

\begin{table}[t]
\caption{Result of pre-analysis}
\label{Result of pre-analysis}
\begin{tabular}{lllllll}\hline
parameter&J0613-0200&J1012+5307&J1600-3053&J1713+0747&J1744-1134&J1909-3744\\\hline
EFAC noise\\
Rcvr\_800\_GASP&${1.096}_{-0.059}^{+0.059}$&${1.138}_{-0.042}^{+0.049}$&${1.19}_{-0.11}^{+0.11}$&${1.134}_{-0.054}^{+0.050}$&${1.168}_{-0.064}^{+0.073}$&$0.984_{-0.060}^{+0.081}$\\
Rcvr\_800\_GUPPI&${1.164}_{-0.032}^{+0.041}$&${1.182}_{-0.023}^{+0.023}$&${1.117}_{-0.040}^{+0.025}$&${1.066}_{-0.020}^{+0.023}$&${1.070}_{-0.024}^{+0.027}$&$1.041_{-0.019}^{+0.023}$\\
Rcvr1\_2\_GASP&${1.061}_{-0.053}^{+0.056}$&${1.054}_{-0.041}^{+0.044}$&${1.155}_{-0.148}^{+0.096}$&${1.087}_{-0.060}^{+0.056}$&${0.991}_{-0.065}^{+0.080}$&$0.977_{-0.060}^{+0.054}$\\
Rcvr1\_2\_GUPPI&${1.080}_{-0.023}^{+0.023}$&${1.081}_{-0.018}^{+0.021}$&${1.062}_{-0.019}^{+0.018}$&${1.040}_{-0.014}^{+0.015}$&${1.079}_{-0.026}^{+0.028}$&$1.049_{-0.013}^{+0.017}$\\
L-wide\_ASP&&&&${1.013}_{-0.048}^{+0.062}$&&\\
L-wide\_PUPPI&&&&${1.091}_{-0.021}^{+0.026}$&&\\
S-wide\_ASP&&&&${1.114}_{-0.058}^{+0.066}$&&\\
S-wide\_PUPPI&&&&${1.110}_{-0.034}^{+0.038}$&&\\\hline
EQUAD noise\footnote{The values are expressed in logarithm with base 10.}\\
Rcvr\_800\_GASP&${-8.5}_{-1.5}^{+1.7}$&${-6.68}_{-3.22}^{+0.18}$&${-7.7}_{-2.2}^{+1.5}$&${-6.97}_{-2.93}^{+0.18}$&${-6.469}_{-0.085}^{+0.079}$&$-6.620_{-0.056}^{+0.046}$\\
Rcvr\_800\_GUPPI&${-6.677}_{-0.188}^{+0.099}$&${-6.348}_{-0.059}^{+0.062}$&${-6.45}_{-3.45}^{+0.14}$&${-7.14}_{-2.75}^{+0.14}$&${-6.598}_{-0.045}^{+0.036}$&$-7.344_{-0.119}^{+0.090}$\\
Rcvr1\_2\_GASP&${-9.74}_{-0.17}^{+3.16}$&${-6.51}_{-3.38}^{+0.19}$&${-6.34}_{-3.54}^{+0.17}$&${-7.29}_{-2.59}^{+0.16}$&${-6.382}_{-0.097}^{+0.072}$&$-7.40_{-2.41}^{+0.18}$\\
Rcvr1\_2\_GUPPI&${-9.65}_{-0.27}^{+2.75}$&${-6.50}_{-0.25}^{+0.11}$&${-8.3}_{-1.7}^{+1.2}$&${-8.01}_{-1.93}^{+0.24}$&${-6.681}_{-0.051}^{+0.041}$&$-8.04_{-1.88}^{+0.19}$\\
L-wide\_ASP&&&&${-7.54}_{-0.23}^{+0.11}$&&\\
L-wide\_PUPPI&&&&${-7.89}_{-1.78}^{+0.13}$&&\\
S-wide\_ASP&&&&${-8.46}_{-1.48}^{+0.93}$&&\\
S-wide\_PUPPI&&&&${-7.64}_{-2.24}^{+0.16}$&&\\\hline
ECORR noise$^{a}$\\
Rcvr\_800\_GASP&${-7.50}_{-0.96}^{+0.69}$&${-8.08}_{-0.38}^{+1.37}$&${-8.472}_{+0.019}^{+2.269}$&${-7.45}_{-1.02}^{+0.49}$&${-6.63}_{-1.38}^{+0.25}$&$-7.95_{-0.52}^{+0.76}$\\
Rcvr\_800\_GUPPI&${-6.74}_{-0.32}^{+0.26}$&${-8.480}_{0.010}^{+1.517}$&${-6.20}_{-0.24}^{+0.13}$&${-6.609}_{-0.142}^{+0.092}$&${-6.40}_{-0.17}^{+0.12}$&$-7.19_{-0.24}^{+0.16}$\\
Rcvr1\_2\_GASP&${-7.94}_{-0.51}^{+1.43}$&${-6.59}_{-1.83}^{+0.28}$&${-6.74}_{-1.70}^{+0.31}$&${-7.11}_{-0.54}^{+0.25}$&${-6.16}_{-0.20}^{+0.14}$&$-8.31_{-0.17}^{+0.89}$\\
Rcvr1\_2\_GUPPI&${-6.77}_{-1.64}^{+0.15}$&${-6.55}_{-1.45}^{+0.15}$&${-6.81}_{-0.34}^{+0.16}$&${-7.115}_{-0.091}^{+0.067}$&${-6.372}_{-0.087}^{+0.090}$&$-7.113_{-0.079}^{+0.066}$\\
L-wide\_ASP&&&&${-6.97}_{-0.11}^{+0.14}$&&\\
L-wide\_PUPPI&&&&${-7.061}_{-0.069}^{+0.087}$&&\\
S-wide\_ASP&&&&${-6.95}_{-0.18}^{+0.14}$&&\\
S-wide\_PUPPI&&&&${-7.03}_{-0.10}^{+0.11}$&&\\\hline
red noise\\
${A}^{a}$&${-13.15}_{-0.68}^{+0.17}$&${-12.67}_{-0.16}^{+0.14}$&${-13.45}_{-6.36}^{+0.16}$&${-14.46}_{-3.74}^{+0.39}$&${-13.44}_{-3.75}^{+0.33}$&$-14.13_{-1.96}^{+0.35}$\\
$\gamma$&${1.26}_{-0.99}^{+2.08}$&${0.96}_{-0.61}^{+0.79}$&${0.17}_{-0.10}^{+6.49}$&${2.3}_{-1.6}^{+3.9}$&${2.9}_{-1.6}^{+3.9}$&$6.899_{-5.295}^{-0.021}$\\
\hline
\end{tabular}
\end{table}

\section{Result}\label{sec:Results}
In this section we describe the upper limits on the amplitude of the FDM signal and how much the FDM signal is absorbed by other noises.
All our result was calculated using six pulsars: PSRs J0613-0200, J1012+5307, J1600-3053, J1713+0747, J1744-1134, and J1909-3744 in the NANOGrav 11-year data set.

\subsection{Upper limits}\label{subsec:Upper limits}
We calculated the 95\% confidence upper limits on the amplitude of the FDM signal by the Bayesian analysis.
We ran all the MCMC simulation with $10^{6}$ iterations and removed the first 25\% as a burn-in period.
As the prior probability distribution of the amplitude of the FDM signal, we considered two cases: the uniform distribution and the log-uniform distribution.
The uniform prior was used to place the conservative upper limits, the log-uniform prior was used to calculate the Bayes factors, where the upper limits were calculated using the \eref{upper limit} and the Bayes factors were calculated using \eref{SDDR}.
In order to see the effect of including the SSE noise in the model on the results, we also calculated the upper limits and the Bayes factors when the SSE noise is not included in the model.

\begin{figure}[t]
\begin{center}
\center
\includegraphics[bb=0 0 611 442, width=13cm]{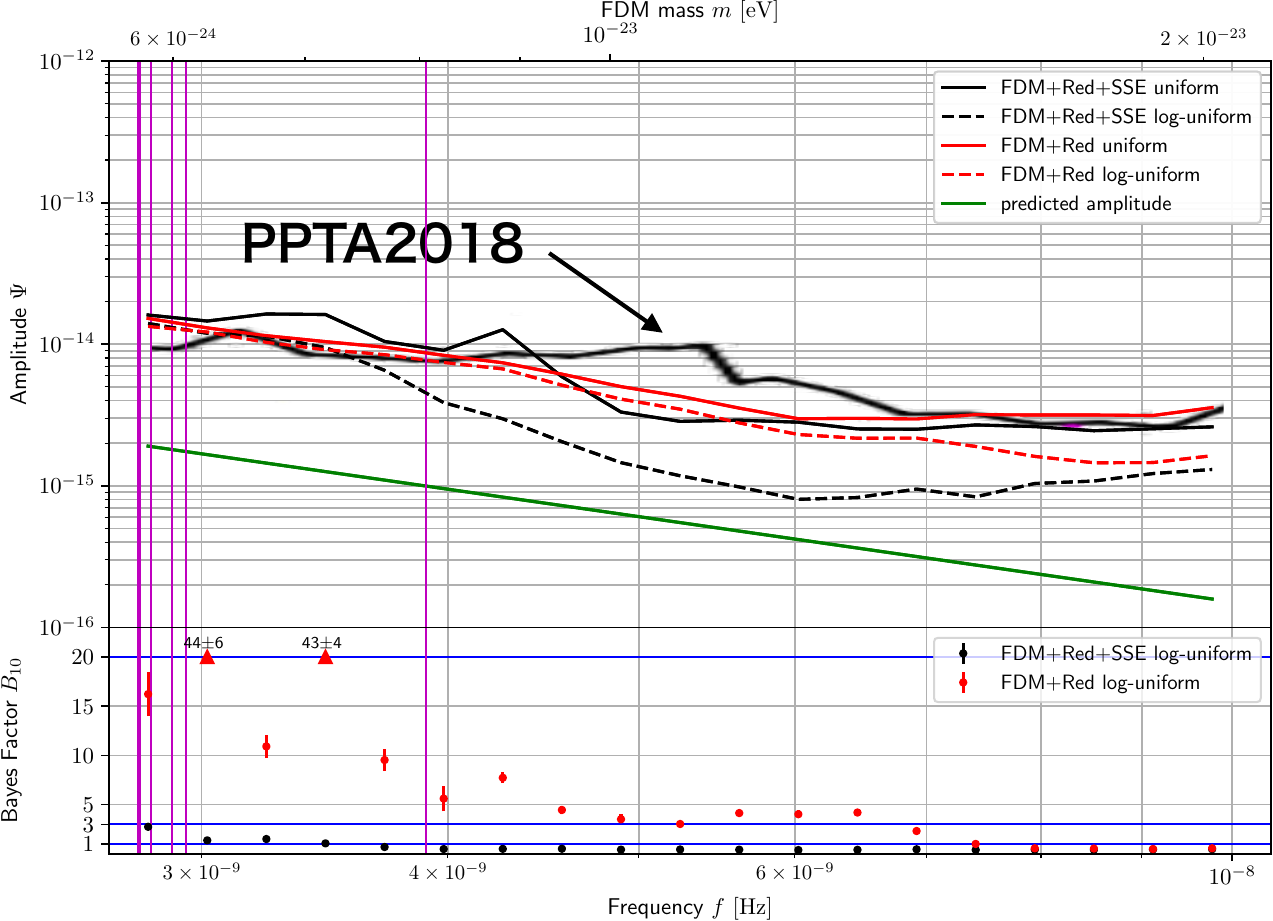}
\caption{Top: The 95\% upper limits on the amplitude of the FDM signal $\Psi$ using the NANOGrav 11-year data set. 
As a prior probability distribution of the amplitude, the uniform prior was used for the black solid lines and the log-uniform prior was used for black dashed lines.
The red lines are the upper limit obtained when the SSE noise is not included in the model describing the observed data, and the solid and dashed lines indicate that unifrom and loguniform were used, respectively.
The bold black line is the upper limit obtained by the Bayesian analysis of the PPTA data ( taken from Figure 3 in  \cite{Porayko:2018sfa}).
The green line is the predicted amplitude.
The purple vertical line is the inverse of the observation period of the pulsars.
Bottom: The values of the Bayes factor obtained when using log-uniform prior.
The black and red indicate when the SSE noise is included in the model or not, respectively.
To improve the visibility of the plot, when the value of the Bayes factor exceeds 20, the upper triangle is used.
 }
\label{Upper limits plot}
\end{center}
\end{figure}

In Figure \ref{Upper limits plot}, we show the upper limits and the Bayes factors of the amplitude $\Psi$ as a function of the frequency $f$ and the mass $m$.
The relation between the frequency $f$ and the mass $m$ is given by \eref{f}.
First, in the above plot, the black solid and dashed lines denote the upper limit using uniform prior and log-uniform prior, respectively.
Here, we plotted the results obtained using log-uniform prior, but as we mentioned in the section \ref{sec:Prior Probability Distribution}, we regard the results obtained by the uniform prior as conservative upper limits.
The red solid and dashed lines denote the upper limit obtained when the SSE noise was not included in the model, and using uniform-prior and log-uniform prior, respectively.
The bold black line denotes the upper limit of the Bayesian analysis obtained in \cite{Porayko:2018sfa} ( taken from Figure 3).
The green line denotes the predicted amplitude of the FDM signal given by \eref{Psi} with $\rho=0.4 {\rm Gev/cm^3}$.
The purple vertical lines denote the inverse of the observation periods of pulsars and corresponds to PSRs J1744-1134, J1012+5307, J1909-3744, J1713+0747, J0613-0200, and J1600-3053 in order from the left.
Next, in the bottom plot, the black and red dots denote the mean value of the Bayes factor using the model with and without the SSE noise, respectively.
The unbiased standard deviation is used for error bars.
Only to make the plot easier to see, when the Bayes factor exceeds 20, it is represented by the upper triangle and the mean value and the unbiased standard deviations of the Bayes factor is written above it.

First we consider the red results obtained when the SSE noise is not included in the model.
It turns out that the upper limits for the log-uniform prior gives stronger limits than for the uniform prior, but the difference is small.
The reason the difference is small is that the Bayes factor exceeds 3 when the frequency becomes $10^{-8.19} \,{\rm Hz}$ ($1.34\times10^{-23}\,{\rm eV}$) or lower.
According to the Table \ref{jeffreys scale}, the Bayes factor exceeds 3 means that there is a signal that is somewhat similar to the FDM signal.
Therefore, whichever prior probability distribution is used, the value of the posterior probability distribution tends to be large at the parameter values of that signal, and as a result the upper limits does not change so much.
From the above, it was found that the FDM signal is positively supported in the wide frequency range. 
However, from a physical point of view, it is hard to think that FDM signals has been found, because the upper limit obtained is about an order of magnitude greater than the expected amplitude and also FDM signals would have only a certain frequency.

Next, we consider the black result obtained when the SSE noise is included in the model.
This result is our main result.
Compared with the case where SSE noise is included in the model, it can be seen that the upper limits of the amplitude of the FDM signal obtained using uniform distribution does not change much.
However, the difference between the upper limits obtained from the different prior probability distributions is large.
The reason for this is that all Bayes factors are smaller than 3 and in most cases they do not exceed 1.
Therefore, in this case we conclude that no FDM signal has been detected.
The red and black results show that the Bayes factor is smaller when the SSE noise is considered.
Thus, we conclude that the SSE noise can mimic the FDM signal.
This result would be inferred from the result that the SSE noise can mimic the stochastic gravitational wave background \cite{Arzoumanian:2018saf}.
In comparison with the published Bayesian upper limits of the amplitude of the FDM signal using the PPTA 12-year data set \cite{Porayko:2018sfa},  i.e. comparing the black and the bold black lines, we found that stronger upper limits were obtained when the frequency was in the range from $10^{-8.34}$ to $10^{-8.19}{\rm Hz}$ (from $9.45\times10^{-24}$ to $1.34\times10^{-23}\,{\rm eV}$).
In this range, up to three times stronger upper limits were obtained, and in other region, about the same upper limits were obtained.

\begin{figure}[t]
\begin{center}
\center
\includegraphics[bb=0 0 611 442, width=13cm]{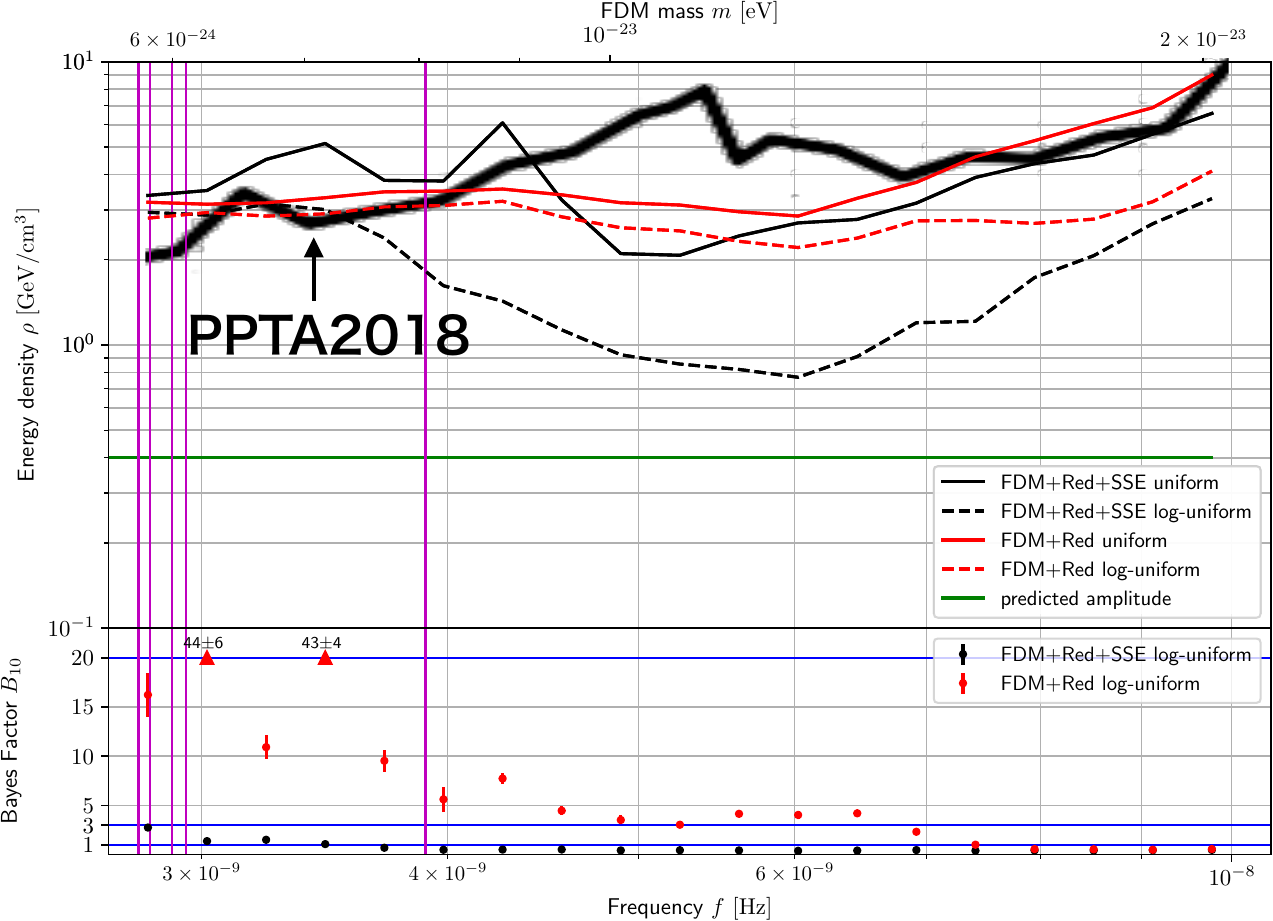}
\caption{Top: The 95\% upper limits on the energy density of the FDM $\rho$ using the NANOGrav 11-year data set. 
This plot is the same as in Figure \ref{Upper limits plot} except that the amplitude is converted to energy density.
The bold black line denotes the upper limit obtained by the Bayesian analysis of the PPTA data ( taken from Figure 4 in  \cite{Porayko:2018sfa}).
Bottom: The values of the Bayes factor obtained when using log-uniform prior.
This plot is the same as in Figure \ref{Upper limits plot}.
 }
\label{Upper limits rho plot}
\end{center}
\end{figure}

It is also important to see the upper limit on the energy density of the dark matter near the Earth rather than the amplitude of the FDM signal.
Thus, we convert the amplitude of the FDM signal into the energy density using \eref{Psi}, and the result is plotted in Figure \ref{Upper limits rho plot}.
Note that the bold black line denotes the upper limit on the energy density with the Bayesian analysis in \cite{Porayko:2018sfa} (taken from Figure 4).
As we can see from Figure \ref{Upper limits rho plot}, our main upper limit represented by the black line is 7 or less in the range from $10^{-8.55}$ to $10^{-8.01}\,{ \rm Hz}$ (from $5.83\times10^{-24}$ to $2.02\times10^{-23}\,{\rm eV}$) where we analyzed.
The strongest upper limit on the the energy density is $2\,{\rm GeV/cm^3}$ at the frequency  $10^{-8.28}\,{ \rm Hz}$ ($1.09\times10^{-23}\,{\rm eV}$).

\subsection{Fixed noise analysis}\label{sec:Fixed noise analysis}
We analyzed the red noise and the SSE noise first and calculated the upper limits of the amplitude of the FDM signal using the obtained MAP values of the parameters.
We ran the MCMC simulation with $10^{6}$ iterations for analysis of red noise and SSE noise and with $10^{5}$ iterations for analysis of the FDM signal, and in both cases we removed the first 25\% as a burn-in period.
The result of the red noise and the SSE noise used for the fixed noise analysis are summarized in Table \ref{red noise and SSE noise used for fixed noise analysis}.
The obtained posterior distribution is expressed by the MAP value and the 95\% confidence interval.

\begin{figure}[t]
\begin{center}
\center
\includegraphics[bb=0 0  611 442, width=13cm]{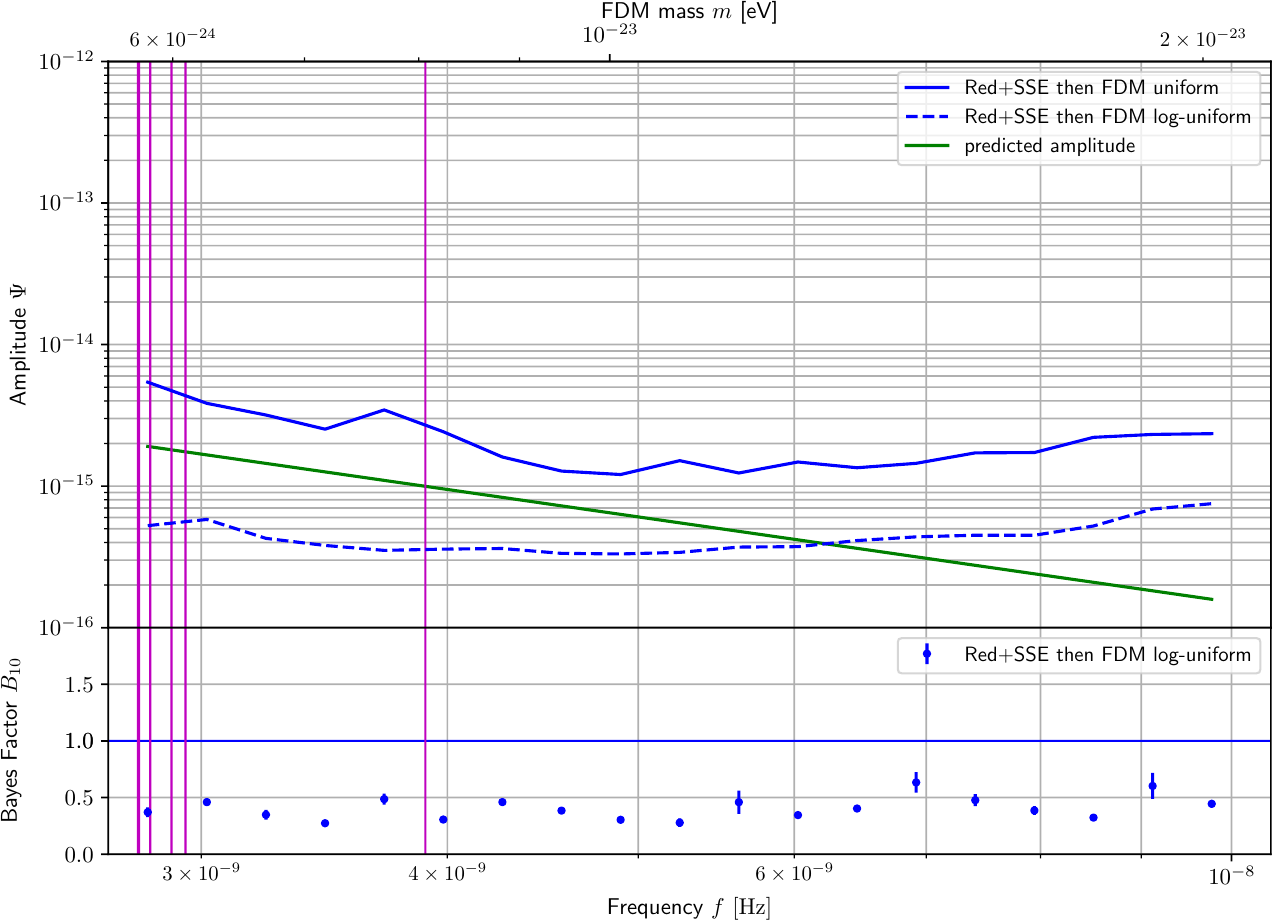}
\caption{Similar plot to Figure \ref{Upper limits plot}.
The FDM signal, the red noise and the noise induced by Jupiter in the SSE noise have similar waveforms and any of these can not be analyzed in advance.
Therefore we do not regard this plot from such an analysis as the upper limits of the amplitude of the FDM signal.
The purpose of this plot is to know how much the red noise and the SSE noise can absorb the signal of the FDM signal.}
\label{Upper limits plot degeneracy}
\end{center}
\end{figure}

As in the previous section, we calculated two cases of uniform and log-unifrom distributions as the prior probability distribution of the amplitude of the FDM signal.
The results are plotted in Figure \ref{Upper limits plot degeneracy} which is the similar plot as Figure \ref{Upper limits plot}.
The solid and dashed lines indicate that the unifrom and the log-uniform prior were used, respectively.

\begin{table}[t]
\caption{Red noise and SSE noise used for fixed noise analysis}
\label{red noise and SSE noise used for fixed noise analysis}
\begin{tabular}{lllllll}\hline
parameter&J0613-0200&J1012+5307&J1600-3053&J1713+0747&J1744-1134&J1909-3744\\\hline
red noise\\
${A_{\rm red}}\footnote{The values are expressed in logarithm with base 10.}$&${-13.20}_{-0.73}^{+0.20}$&${-12.68}_{-0.16}^{+0.14}$&${-13.51}_{-6.34}^{+0.10}$&${-19.939}_{0.055}^{+5.436}$&${-13.51}_{-2.31}^{+0.33}$&${-14.05}_{-5.77}^{+0.17}$\\
$\gamma$&${1.36}_{-0.98}^{+2.33}$&${1.07}_{-0.63}^{+0.74}$&${0.0897}_{-0.0069}^{+6.6316}$&${1.8}_{-1.6}^{+4.9}$&${2.8}_{-1.2}^{+4.0}$&${0.74}_{-0.62}^{+5.97}$\\
\hline
&\multicolumn{6}{c}{common to all pulsars}\\
\hline
SSE noise\\
$\delta M^{\rm J}$&\multicolumn{6}{c}{${-2\times10^{-12}}_{-3.0\times10^{-11}}^{+3.3\times10^{-11}}$}\\
$\delta M^{\rm S}$&\multicolumn{6}{c}{${1\times10^{-12}}_{-1.6\times10^{-11}}^{+1.6\times10^{-11}}$}\\
$\delta M^{\rm U}$&\multicolumn{6}{c}{${1\times10^{-11}}_{-1.2\times10^{-10}}^{+1.0\times10^{-10}}$}\\
$\delta M^{\rm N}$&\multicolumn{6}{c}{${1\times10^{-11}}_{-1.7\times10^{-10}}^{+1.5\times10^{-10}}$}\\
$\delta a_{1}^{\rm J}$&\multicolumn{6}{c}{${-0.0046}_{-0.0099}^{+0.0109}$}\\
$\delta a_{2}^{\rm J}$&\multicolumn{6}{c}{${0.004}_{-0.018}^{+0.019}$}\\
$\delta a_{3}^{\rm J}$&\multicolumn{6}{c}{${-0.0128}_{-0.0083}^{+0.0071}$}\\
$\delta a_{4}^{\rm J}$&\multicolumn{6}{c}{${-0.0123}_{-0.0077}^{+0.0117}$}\\
$\delta a_{5}^{\rm J}$&\multicolumn{6}{c}{${0.0044}_{-0.0087}^{+0.0080}$}\\
$\delta a_{6}^{\rm J}$&\multicolumn{6}{c}{${0.008}_{-0.019}^{+0.012}$}\\
$\delta z$&\multicolumn{6}{c}{${-2.0\times10^{-10}}_{-7.4\times10^{-10}}^{+1.15\times10^{-09}}$}\\
\hline
\end{tabular}
\end{table}

The reason for doing this analysis is to know how much the red noise and the SSE noise can absorb the signal of the FDM signal. 
Note that this analysis is not intended to give the upper limits on the amplitude of the FDM signal.
The FDM signal, the red noise and the noise induced by Jupiter in the SSE noise have similar waveforms and any of these can not be analyzed in advance.
We consider that this analysis is not suitable for giving an upper limit to the amplitude of the FDM signal, and the obtained results are not regarded as the upper limits of the amplitude of the FDM signal.

It can be seen from Figure \ref{Upper limits plot degeneracy} that the values of the upper limits are drastically smaller than the result obtained in the previous section.
In particular, when using the log-uniform prior,  surprisingly, the upper limits are smaller than the predicted amplitude in some range.
As for Bayesian factors, they are all smaller than 1, which is consistent with the fact that the upper limits are strongly influenced by the prior probability distribution.
From this result, it is inferred that the FDM signal is well absorbed by the red noise and the SSE noise.

In order to investigate the impact of analyzing the noise first, we made simulated noise using MAP values of the SSE noise
Then we calculated the Lomb-Scargle periodogram of the timing residual by subtracting it, where the Lomb-Scargle periodogram can be used to search for periodic signals in non-uniformly spaced time series data \cite{1976Ap&SS..39..447L,1982ApJ...263..835S}.
The reason for not subtracting the red noise from the original timing residual is that it is difficult to create the noise included in the actual data because the red noise that we can create is only one realization of the stochastic process.
For comparison, we also calculated the Lomb-Scargle periodogram of the original timing residual and the simulated timing residual induced by the red noise only.
In order to calculate the Lomb-Scargle periodogram, we used Astropy\footnote{http://www.astropy.org} \cite{2013A&A...558A..33A, 2018AJ....156..123A} which is a Python package for the astronomy.
For the purpose of expressing red noise, the original residual has short observation period and lacks frequency resolution, so that we made red noise using simulated observation data that the observation period is $10^5$ days and the data points are every day.

\begin{figure}[t]
\begin{center}
\center
\includegraphics[bb=0 0 720 720, width=18cm]{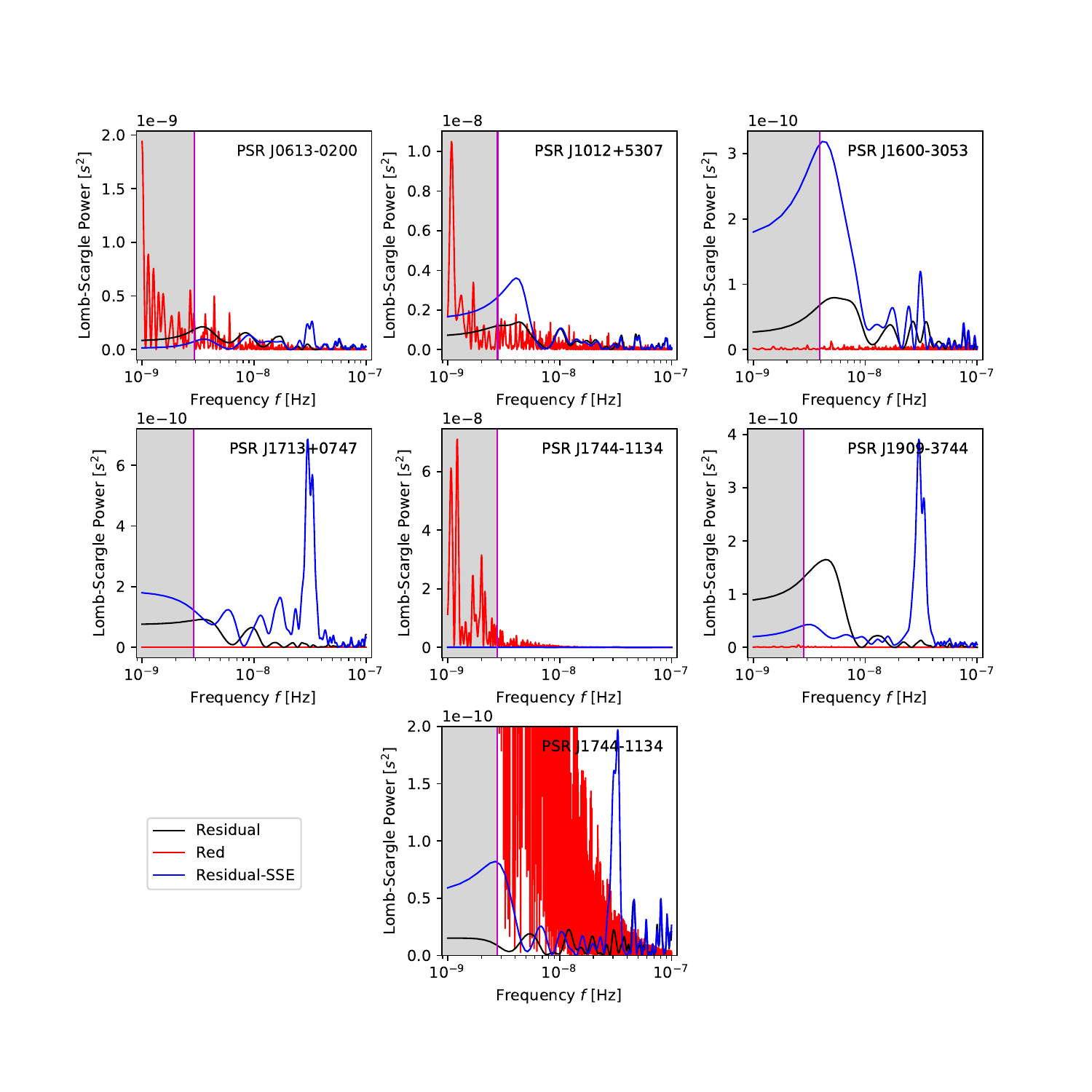}
\caption{The Lomb-Scargle periodogram for PSRs J0613-0200, J1012+5307, J1600-3053, J1713+0747, J1744-1134, and J1909-3744.
The black, red, and blue lines denote the Lomb-Scargle periodogram of the original timing residual, only the red noise, and the original timing residual after subtracting the SSE noise.
As for PSR J1744-1134, the red noise is large, hence a figure with a different scale  is depicted below that figure to see  the other lines.
 }
\label{subtract_red+ephem}
\end{center}
\end{figure}

In Figure \ref{subtract_red+ephem}, we show the Lomb-Scargle periodogram, where the horizontal axis is the frequency and the vertical axis is the Lomb-Scargle power.
The black, red, and blue lines represent the periodogram of the original timing residual, the red noise only, and the timing residual subtracted by the SSE noise, respectively.
As for PSR J1744-1134, a plot focused on other than the red noise is also displayed below it, because the red noise is large and the other lines are difficult to see.
The purple vertical line represents the inverse of the observation period.

First, with regard to the black line representing the periodogram of the original timing residual, it can be seen that the power in the frequency smaller than $10^{-8} {\rm Hz}$ is larger than the power in the frequency larger than $10^{-8} {\rm Hz}$.
Therefore, it can be understood that there is noise in the low frequency region, and the analysis of this article is considered to be meaningful.

Second, with regard to the red line representing the periodogram of the red noise only, the red noise can be considered to characterize low frequency noise in PSRs J0613-0200 and J1012+5307, but it can not be considered so in the other pulsars.
In particular, the amplitude of red noise is large in PSR J1744-1134, so that in the analysis in which the MAP value of the red noise is fixed, it can be seen that this pulsar will contribute little to the analysis result of the amplitude of the FDM signal.
The reason why the red noise can be detected properly in PSRs J0613-0200 and J1012+5307 is that, according to the result of analyzing only the red noise and the SSE noise, the posterior distribution of the red noise is obtained with a sharp peak.
On the other hand, in the other pulsars, the posterior probability distribution of the red noise has a peak which spreads over a wide parameter space, which indicates that the MAP value has little meaning.

Finally, with regard to the blue line representing the original timing residual after subtracting the SSE noise, it is found that the low frequency noise are reduced in PSRs J0613-0200 and J1909-3744.
On the other hand, this is not the case with the other pulsars, and it can be seen that the noise increases rather than decreases.
We conclude that the reason why the upper limits are smaller in fixed noise analysis than in the main analysis is mainly due to PSR J1909-3744, because only this pulsar has small red noise and the low frequency noise is removed by the SSE noise.
Conversely, the other pulsars have not improved much, which may suggest that fixed noise analysis is not successful.
In particular, since the SSE noise can mimic the FDM signal, it is a problem that the model of the SSE noise may favor parameters which are contrary to the expectation that noise will be reduced on all pulsars.
More precise measurements of the SSE would be needed to detect or exclude the FDM signal.
However, these issues are beyond the scope of this article.

\section{Conclusion}
We searched for the FDM by performing the Bayesian analysis in the time domain using the NANOGrav 11-year Data Set.
In Section \ref{subsec:Upper limits}, we gave the 95\% confidence upper limit on the amplitude of the FDM signal. we found that probability that the FDM signal should be included in the model was less than 75\% in all frequency region.
Compared with the published Bayesian upper limit of the FDM signal using the PPTA 12-year data set \cite{Porayko:2018sfa}, we found that our upper limit was up to 3 times stronger than the previous study when the frequency was in the range from $10^{-8.34}$ to $10^{-8.19}{\rm Hz}$ ($9.45\times10^{-24}$ to $1.34\times10^{-23}\,{\rm eV}$ in terms of the FDM mass).
In other region, we also obtained the similar upper limit on the amplitude of the FDM signal.
Since the amplitude of FDM signal can be converted to the energy density of the dark matter near the Earth, it is easy to obtain the upper limit of the energy density.
The upper limit on the energy density was lower than $7\,{\rm GeV/cm^3}$ in the range from $10^{-8.55}$ to $10^{-8.01}\,{ \rm Hz}$ (from $5.83\times10^{-24}$ to $2.02\times10^{-23}\,{\rm eV}$) 
where we analyze.
In particular, at a frequency of $10^{-8.28}\,{ \rm Hz}$ (a mass of $1.09\times10^{-23}\,{\rm eV}$), we obtained the strongest upper limit  $2\,{\rm GeV/cm^3}$.
In addition to the main analysis, we also investigated the case where the SSE noise was not included in the model.
In this case, we showed that we can not exclude the existence of the FDM signal, because the probability that the FDM signal should be included in the model was more than 75\% in the frequency region $10^{-8.19} \,{\rm Hz}$ or less.
This results show that the Bayes factor is smaller when the SSE noise is considered.
Thus, we conclude that the SSE noise can mimic the FDM signal, which would be inferred from the result that the SSE noise can mimic the stochastic gravitational wave background \cite{Arzoumanian:2018saf}.

In Section \ref{sec:Fixed noise analysis}, by analyzing the noise in advance, we examined how much the FDM signal was absorbed.
In this case, we clarified that the probability that the FDM signal should be included in the model was much lower than 50\% in all frequency region.
Compared to our main analysis, we found that the upper limit on the amplitude of the FDM signal became very small.
From this, it is expected that the FDM signal will be absorbed very well by analyzing the noise in advance.
Then, we made a simulated noise from the parameters obtained by the analysis earlier, and investigated whether removing the SSE noise from the observed data would reduce the power of the low frequency region of the data.
We found that the power of the PSR J1909-3744 has become smaller, and we conclude that this pulsar contributed to the improvement of the sensitivity.
With other pulsars, we also found that the power increased on the contrary.
If the SSE noise is properly found, it is thought that the noise will be reduced for all pulsars.
Hence, this result seems strange.
In order to solve this problem, it is necessary to reduce the uncertainty of the SSE by more accurate observation of the SSE.

\acknowledgments
R.K. was supported
by Grant-in-Aid for JSPS Research Fellow and JSPS KAKENHI Grant Numbers
17J00496.
J.~S. was in part supported by JSPS KAKENHI Grant Numbers
JP17H02894, JP17K18778, JP15H05895, JP17H06359, JP18H04589. J.~S
is also supported by JSPS Bilateral Joint Research Projects (JSPS-NRF
collaboration) “String Axion Cosmology.”

\bibliography{CambridgeCore_CitationExport_22Jul2018,Doing_Bayesian_Data_Analysis,Bayesian_Data_Analysis_Third_Edition,The_Theory_of_Probability,tandf_uasa2090_773,Kendall_s_advanced_theory_of_statistics,Bayesian_Model_Selection_in_Social_Research,10_2307_2958475,tandf_uasa2090_614,nano11sto,3101501citation,tempo2II,nano9_result,nano11_result,Boynton_et_al_1972,Groth_1975,shannon_&_codes_2010,Lyne_et_al_2010,Lam_et_al_2017,Shannon_et_al_2013,Lam_et_al_2016,Jenet_et_al_2006,tempo2III,Lentati_et_al_2013,Arzoumanian_et_al_2014,Ellis_2014,Hellings_downs_1983,Champion_et_al_2010,Brouwer_Clemence_1961,Lazio_2018,Blandford_1984,Cutler_et_al_2013,Luzum_et_al_2011.bib,Folkner_et_al_2014,Folkner_et_al_2016,Taylar_et_al_2016,Jacobson_2014.bib,justin_ellis_2017_1037579,Haario_et_al_2001,Haario_et_al_2005,Braak_et_al_2006,nano9sto,PAL2,Porayko_et_al_2018,astropy_2013,astropy_2018,Hu_et_al_2000,Khmelnitsky_et_al_2013,Detweiler_1979,Romani_1989,Foster_et_al_1990,Kramer_et_al_2013,McLaughlin_et_al_2013,manchester_et_al_2013,Hobbs_et_al_2010,Khmelnitsky_Rubakov_2014,Porayko_Postnov_2014,Nesti_et_al_2013,Lomb_1976,Scargle_1982,GWdetectors,Wang_2017}

\end{document}